\documentclass[aps,prd,twocolumn,groupedaddress,showpacs,nofootinbib,amssymb]{revtex4-1}
\usepackage[T1]{fontenc}
\usepackage[latin1]{inputenc}
\usepackage{graphicx}
\usepackage[english]{babel}
\usepackage{amsmath}
\usepackage{mathrsfs}
\usepackage{amssymb}
\usepackage{amsfonts}
\usepackage{epsfig}
\usepackage{colordvi}
\usepackage{color}
\usepackage{bm}

\def\de#1/de#2{\frac{\partial {#1}}{\partial {#2}}}
\def\De#1/de#2{\dfrac{\partial {#1}}{\partial {#2}}}

\begin{document}
\title{Running coupling in electroweak interactions of leptons from $f(R)$-gravity with torsion}
\author{Salvatore Capozziello$^{1,2}$, Mariafelicia De Laurentis$^{1,2}$, Luca Fabbri$^{3,4,5}$, Stefano Vignolo$^{3}$}
\affiliation{\it $^1$Dipartimento di Scienze Fisiche, Universit\`{a} di Napoli ``Federico II'', Compl. Univ. di Monte S. Angelo, Edificio G, Via Cinthia, I-80126, Napoli, Italy\\
$^2$ INFN Sez. di Napoli, Compl. Univ. di Monte S. Angelo, Edificio G, Via Cinthia, I-80126, Napoli, Italy\\
$^3$DIPTEM Sez. Metodi e Modelli Matematici, Universit\`{a} di Genova, Piazzale Kennedy, Pad. D, 16129 Genova, Italy\\
$^4$INFN Sez. di Bologna, Viale B. Pichat 6/2, I-40127 Bologna, Italy\\
$^5$ Dipartimento di Fisica, Universit\`{a} di Bologna, Via Irnerio 46, I-40126 Bologna, Italy}
\date{\today}
\begin{abstract}
The $f(R)$-gravitational theory with torsion is considered for one family of leptons; it is found that the torsion tensor gives rise to interactions having the structure of the weak forces while the intrinsic non-linearity of the $f(R)$ function provides an energy-dependent coupling: in this way, torsional $f(R)$ gravity naturally generates both structure and strength of the electroweak interactions among leptons. This implies that the weak interactions among the lepton fields could be addressed as a geometric effect due to the interactions among spinors induced by the presence of torsion in the most general $f(R)$ gravity. Phenomenological considerations are addressed.
\end{abstract}
\maketitle
\section{Introduction}
\label{Intro}
So far back as forty years ago, all the elementary particles that were known at the time have been placed into a schematic and unique frame, thus called the Weinberg-Salam Standard Model (SM); such SM has further gained much consensus due to the fact that the predictions it was able to make have been experimentally measured to a very high precision level.

The fundamental structure of the SM is summarized in its Lagrangian $\mathscr{L}_{\rm{SM}}$, which can be written in the general form as
\begin{equation}
\label{LagrangianSM}
\mathscr{L}_{\rm{SM}}=
\mathscr{L}_{\rm{Dirac}}+\mathscr{L}_{SU(2)_{L}\times U(1)}+\mathscr{L}_{\rm{Higgs}}\,.
\end{equation}
The term $\mathscr{L}_{\rm{Dirac}}$ describes dynamics of Dirac massless spinors; further it is assumed that these massless spinors have their left-hand projections mixing according to the $SU(2)_{L}$ isospinorial symmetry beyond the $U(1)$ hypercharge symmetry, so the term $\mathscr{L}_{SU(2)_{L} \times U(1)}$ is added to describe dynamics of the complete $SU(2)_{L}\times U(1)$ gauge interactions, giving rise to massless interactions; however, in nature, we observed interactions mediated by massive vector bosons, and then it is assumed that these gauge massless fields lose their gauge symmetry while getting masses, and the term $\mathscr{L}_{\rm{Higgs}}$ is eventually added to describe dynamics of the Higgs scalar field that would induce the gauge symmetry breaking and mass generation mechanism. When in the Lagrangian $\mathscr{L}_{\rm{SM}}$, the Higgs field is in a specific vacuum configuration, the initial $SU(2)_{L}\times U(1)$ gauge symmetry is broken and all fields coupled to the Higgs acquire their masses; in order for the Lagrangian $\mathscr{L}_{\rm{SM}}$ to adequately represent the elementary particles observed in nature, the Higgs terms $\mathscr{L}_{\rm{Higgs}}$ is necessary. Albeit within the Lagrangian $\mathscr{L}_{\rm{SM}}$, the fermionic and interacting terms have been experimentally confirmed, there is for the Higgs term no evidence yet and the detection of such a field can be considered as one of the biggest challenges for LHC experiments now running at CERN.

Clearly, the fact that the Higgs field has not been observed at the moment does not mean it could not be observed in the future; however, its discovery was supposed to be within the reach of present accelerators: as this situation persists as reported by ATLAS collaboration at CERN \cite{ATLAS} and CDF collaboration at Fermi Lab \cite{CDF}, it becomes more and more compelling the possibility that the Higgs field may not even exist. On the other hand however, if the Higgs field does not exist then also the predicted phenomenology cannot be obtained after gauge symmetry breaking and mass generation from the Lagrangian $\mathscr{L}_{\rm{SM}}$: therefore if within the Lagrangian $\mathscr{L}_{\rm{SM}}$ there is no $\mathscr{L}_{\rm{Higgs}}$ term then no breaking of the symmetry for the $SU(2)_{L}\times U(1)$ gauge group can be achieved and furthermore no generation of the masses of the elementary particles can be possibly accomplished; since the $SU(2)_{L}\times U(1)$ gauge group is allowed precisely because the spinors are massless then to suppress the existence of at least the $SU(2)_{L}$ sector of the gauge group, we have to require the spinors to be massive from the very beginning. This requirement however means that the mass of elementary particles has to be present at any energy scale generating a dramatic hierarchy problem \cite{hierarchy}.

Because of the presence of the mass terms there can be no gauge symmetry, and therefore there can be no fundamental $W^{\pm}$ and $Z$ boson, so that whatever will play the role of the $W^{\pm}$ and $Z$ bosons must be a composite vector field, yet able to recover the phenomenology of the fermionic weak forces we have already observed, as long as the energy is low enough not to probe their compositeness; and this model would have no Higgs boson, whether fundamental or composite. Now the question we have to ask is: what could give rise to $W^{\pm}$ and $Z$ bosons with composite structure, and no Higgs field at all?

The answer to this question has to find place in a new type of physics that could eventually involve the gravitational sector. Actually, there is a type of physics that is relatively new in its applications to the physics of elementary particles because, in the past, people almost never thought to employ it to study high-energy systems: this type of physics might well be based on the Cartan theory of torsion \cite{cartan} and the so-called {\it Extended Theories of Gravity} (ETGs) \cite{book,physrep,physOdi} involving further degrees of freedom for the gravitational field.

As we shall better discuss in the following, Cartan torsion tensor is indeed a quantity that is an integral part of the most general connection in the General Theory of Relativity, and it is actually capable of introducing new dynamical effects by endowing with non-linear potentials the fermionic field equation; these potentials are intriguingly similar to the weak forces of the Fermi Model as it has been remarked since the beginning of the 1970s in the seminal works \cite{h-d,h-h-k-n,s-s,ds-g,ds-s} and references therein. That torsion induces electroweak-like interactions not only at the low-energy scale of the effective Fermi Model but also up to the higher-energy scale of the SM has been shown in \cite{f/1,f/2}; eventually, a model in which the torsion induces electroweak interactions not as gauge fields before the symmetry breaking but directly as effective weak forces has been developed in \cite{F/1}: all results exposed in \cite{h-d,s-s,f/1,f/2,F/1} collected together show that by starting from a theory of gravity in which both metric and torsional degrees of freedom are considered, it is possible to build a model for the system of two fermions. Torsion induces fermions to couple as if they were actually weakly coupled, as experimentally observed \cite{shapiro}. More importantly, in the present construction, the weak coupling looks as if mediated by composite massive bosons whereas in the SM they are supposed to be elementary; this means that in these approaches the $W^{\pm}$ and $Z$ bosons have an internal structure they should display at high-energy scales, whereas in the SM they must be structureless at any energy scale. But even more importantly, there is here no hint of any Higgs boson, whatever composite or fundamental. All this is certainly intriguing, but, in order for the entire idea to be acceptable, there is a fundamental obstacle that must first be overcome: it concerns the fact that torsion should become relevant not little before or at the Fermi or Electroweak scale but well beyond it. The mechanism by which to achieve such a result could be provided by ETGs which naturally induce a gravitational running coupling, as it has been shown in \cite{CBD}.

Indeed, in all models in which torsion is thought to give rise to electroweak interactions or weak forces, a mechanism for an energy-dependent running of the gravitational coupling must take place, in order for the gravitational effects to reach the Fermi scale while starting from the Planck scale. Interestingly, this mechanism might come from a straightforward generalization of the gravitational interaction, e.g. $f(R)$ gravity, because it concerns a Lagrangian function that is not simply linear in Ricci scalar $R$ but a generic function of it as discussed in \cite{CCSV1,CCSV2,CCSV3,CV4} and in more general instances in \cite{book}. This theory, and in general ETGs, has recently acquired great interest in cosmology and astrophysics in addressing problems related to the dark sector of the matter-energy content of the universe \cite{odintsov,francaviglia}. In principle, as discussed in \cite{physOdi,troisi}, it is possible to reconstruct all phases of cosmic history by suitable $f(R)$ models. This means that inflation, radiation dominated, matter dominated and dark energy phases can be addressed without adding new exotic ingredients but just supposing that the effective theory of gravity can be scale dependent. On the other hand, the same philosophy works for self-gravitating systems: in \cite{cardone,salzano,napolitano}, it is shown that dynamics of low surface brightness spiral galaxies, clusters of galaxies and elliptical galaxies can be matched by corrections to the Newtonian potential coming from $f(R)$ gravity. In particular, Taylor expansions of analytic $f(R)$ models give rise to Yukawa corrections to the gravitational potential that fix the characteristic scales of such astrophysical systems without dark matter. Furthermore, such corrections well evade Solar System tests of General Relativity \cite{tsujikawa} since their effects become relevant from the Galactic scale and beyond. This means that General Relativity works well at local scales but it has to be revised at infrared scales. The point is now to show that a similar paradigm, assuming different couplings, can work also at ultraviolet scales.

Recently $f(R)$ gravity has been considered at high-energy scales in connection to the hierarchy problem and mass generation \cite{CBD} and in presence of Dirac matter fields \cite{f-v}: in particular, by applying $f(R)$ gravity to the specific case of Dirac matter fields, the non-linearity of the function $f(R)$ is equivalent to the presence of a suitable scalar field depending on the bilinear scalar obtained from the Dirac field alone, scaling the potential of the Dirac matter fields within the Dirac matter field equation. Henceforth, if the idea of exploiting torsion to produce electroweak-like forces is performed within the framework of $f(R)$ gravity for the Dirac matter field, then we may expect that all torsion interactions may not only have the form but also the strength of the weak forces at the electroweak energy scale if a suitable fine-tuning is chosen.

The aim of this paper is to show that taking into account $f(R)$ gravity with torsion, it is possible to achieve a natural model for electroweak interactions with a running coupling induced by gravitational degrees of freedom where interactions with fermions are considered. First of all, we shall recall the general properties of torsional $f(R)$ gravity when Dirac matter is present as source; then we shall study the two fermion field model, one of which massless and left-handed; eventually, we are going to show that, for this two-lepton system, the torsional interactions among the two leptons may be written in the form of the weak interactions, mediated by composite $W^{\pm}$ and $Z$ bosons, without Higgs field at all: these torsional weak interactions, mediated by the $W^{\pm}$ and $Z$ bosons, present a coupling constant that runs thanks to the non-linearity of $f(R)$ gravity, and it may eventually be fine-tuned on the value we expect to observe. A phenomenological discussion of gravitational cut-off and electroweak interactions at TeV scales is given.
\section{Dirac matter fields in $f(R)$ gravity}
In this paper, we shall employ both coordinate and tetrad formalisms, the former in Latin and the latter in Greek indices; in the coordinate formalism, Latin indices are lowered and raised by means of the metric $g_{ij}$ and $g^{ij}$, while in the tetrad formalism, Greek indices are lowered and raised by means of the metric $\eta_{\mu\nu}$ and $\eta^{\mu\nu}$ having the Minkowskian form $\mathrm{diag}(1,-1,-1,-1)$: the passage between the two formalism is achieved by means of the tetrad bases $e^{i}_{\mu}$ and $e_{i}^{\mu}$. The differential properties are given in terms of the covariant derivative $\nabla_{i}$, whose commutators define the Riemann curvature tensor $R^{stij}$ and the Cartan torsion tensor $T_{ijh}$; the only contraction of the Riemann curvature tensor defined as $R^{stij}g_{si}=R^{tj}$ is the Ricci curvature tensor, whose contraction $R^{st}g_{st}=R$ is the Ricci curvature scalar, while the contraction of the Cartan torsion tensor $T_{ij}^{\;\;j}=T_{i}$ is the Cartan torsion vector, useful in the following. For all properties of these geometrical objects and their relationships, we refer the reader to references \cite{CCSV1,CCSV2,CV4,f-v}.

The general theory of $f(R)$ gravity possessing both metric and torsional degrees of freedom can be formulated within both the metric-affine and the tetrad-affine framework. According to the basic paradigm of $f(R)$ gravity, the gravitational Lagrangian of the theory is assumed to be a real scalar function $f(R)$, where $R$ is the Ricci curvature scalar written in terms of a metric $g$ and a $g$-compatible connection $\Gamma$ with torsion, or equivalently in terms of a tetrad field $e$ and a spin-connection $\omega$; the pairs $(g,\Gamma)$ or equivalently $(e,\omega)$ represent the gravitational dynamical fields of the theory in the metric-affine and tetrad-affine approaches respectively: by varying the Lagrangian with respect to $(g,\Gamma)$ or $(e,\omega)$ we get in any case the system of field equations for the curvature-energy and torsion-spin couplings
\begin{subequations}
\label{geometricequations}
\begin{eqnarray}
\label{curvature-energy}
&f'(R)R_{ij} -\frac{1}{2}f(R)g_{ij}=\Sigma_{ij}\\
\nonumber
\label{torsion-spin}
&f'(R)T_{ijh}
=\frac{1}{2}\left(g_{ih}\partial_{j}f'(R)-g_{jh}\partial_{i}f'(R)\right)+\\
&+\left(S_{ijh}+\frac{1}{2}g_{ih}S_{jq}^{\;\;\;q}-\frac{1}{2}g_{jh}S_{iq}^{\;\;\;q}\right)
\end{eqnarray}
\end{subequations}
where $\Sigma_{ij}$ and $S_{ijh}$ are the stress-energy and the spin density tensors of matter fields. It is possible to see that from \eqref{torsion-spin} there could be torsion even in absence of the spin density tensor: therefore we may say that while the energy is the source of the curvature of the spacetime, both the non-linearity of $f(R)$ gravity and the spin are the sources of the spacetime torsion \cite{CCSV1,CCSV2,CV4}.

Then, making use of the Bianchi identities, it is possible to derive the conservation laws
\begin{subequations}
\label{conslaws}
\begin{eqnarray}
\label{consenergy}
&\nabla_{i}\Sigma^{ij}+T_{i}\Sigma^{ij}-\Sigma_{pi}T^{jpi}-\frac{1}{2}S_{sti}R^{stij}=0\\
\label{consspin}
&\nabla_{h}S^{ijh}+T_{h}S^{ijh}+\Sigma^{ij}-\Sigma^{ji}=0
\end{eqnarray}
\end{subequations}
which the stress-energy and spin density tensors of matter fields must satisfy, once the matter field equations are assigned. These results have been proven to hold in the most general case \cite{f-v}.

Because in this paper we will deal with $f(R)$ gravity coupled to the Dirac matter field, we need to define the Dirac matrices $\gamma^{\mu}=\gamma^{i}e^{\mu}_{i}$ verifying the Clifford algebra as usual; the commutator $S_{ij}=\frac{1}{8}[\gamma_{i},\gamma_{j}]$ will be used in the following. To extend the covariant derivative to Dirac spinor fields $\psi$ and Dirac conjugate spinor field $\bar{\psi}=\psi^{\dagger}\gamma^{0}$ as $D_{i}\psi=\partial_{i}\psi-\Omega_{i}\psi$ and $D_{i}\bar{\psi}=\partial_{i}\bar{\psi}+\bar{\psi}\Omega_{i}$ we have to define the spinorial connection
\begin{eqnarray}
\label{spinorialconn}
&\Omega_{i}=\frac{1}{4}\omega_{i\;\;}^{\;\mu\nu}\gamma_{\mu}\gamma_{\nu}
=-\frac{1}{4}\left(\Gamma_{ik}^{\;\;\;j}-e^{j}_{\mu}\partial_{i}e^{\mu}_{k}\right)
\gamma_{j}\gamma^{k}
\end{eqnarray}
in terms of the spin connection $\omega_{i\;\;}^{\;\mu\nu}$ or the coordinate connection $\Gamma_{ik}^{\;\;\;j}$ equivalently.

With these geometrical quantities, it is possible to define the stress-energy and spin density tensors
\begin{subequations}
\label{consquantities}
\begin{eqnarray}
\label{energy}
&\Sigma_{ij}=\frac{i}{4}\left(\bar{\psi}\gamma_{i}D_{j}\psi-D_{j}\bar{\psi}\gamma_{i}\psi\right)\\
\label{spin}
&S_{ijh}=\frac{i}{2}\bar{\psi}\left\{\gamma_{h},S_{ij}\right\}\psi
\equiv-\frac{1}{4}\epsilon_{ijhk}\left(\bar{\psi}\gamma_{5}\gamma^{k}\psi\right)
\end{eqnarray}
\end{subequations}
in terms of the parity-odd $\gamma_{5}=i\gamma_{0}\gamma_{1}\gamma_{2}\gamma_{3}$ and because of the presence of the parity-odd completely antisymmetric tensor of Levi--Civita shows that the Dirac spin density tensor is completely antisymmetric as well.

Finally, the Dirac equations are given by
\begin{eqnarray}
\label{matter}
&i\gamma^{k}D_{k}\psi+\frac{i}{2}\gamma^{k}T_{k}\psi-m\psi=0
\end{eqnarray}
in terms of the mass $m$ of the Dirac matter field.

By substituting the energy and spin density tensors in the field equations \eqref{geometricequations} we finally have
\begin{subequations}
\label{geomfieldequations}
\begin{eqnarray}
\label{curv-matterenergy}
&f'(R)R_{ij} -\frac{1}{2}f(R)g_{ij}=\frac{i}{4}\left(\bar{\psi}\gamma_{i}D_{j}\psi-D_{j}\bar{\psi}\gamma_{i}\psi\right)\\
\nonumber
\label{tors-matterspin}
&f'(R)T_{ijh}
=\frac{1}{2}\left(g_{ih}\partial_{j}f'(R)-g_{jh}\partial_{i}f'(R)\right)-\\
&-\frac{1}{4}\epsilon_{ijhk}\left(\bar{\psi}\gamma_{5}\gamma^{k}\psi\right)
\end{eqnarray}
\end{subequations}
linking the Dirac matter to the spacetime geometry.

Of course, once the matter field equations \eqref{matter} are used for the conserved quantities given by the energy and spin density tensors in the right-hand side of the field equations \eqref{geometricequations}, then we have that the conservation laws are verified identically, showing that the theory is consistently defined \cite{f-v}.

In the following, we suppose that the trace of the field equation \eqref{curv-matterenergy}
\begin{equation}
\label{trace}
f'(R)R-2f(R)=\Sigma^{i}_{\;i}=\Sigma
\end{equation}
gives rise to an invertible relation between the Ricci curvature scalar $R$ and the trace of the stress-energy tensor; also, because the case $f(R)=kR^2$ is only compatible with the traceless energy condition, then, for the sake of generality, we assume that $f(R)\not=kR^2$. Under the assumed conditions, from equation \eqref{trace}, it is possible to derive the expression of $R$ as function of $\Sigma$ as $R=F(\Sigma)$ (see \cite{CCSV1,CCSV2,CCSV3,CV4,f-v}). Due to the irreducibility of the Dirac spin density tensor, the torsion tensor has only two irreducible decompositions, given by the completely antisymmetric part, that is
\begin{eqnarray}
\label{dualvector}
&f'(R)T_{ijh}\epsilon^{ijhk}=\frac{3}{2}\left(\bar{\psi}\gamma_{5}\gamma^{k}\psi\right)
\end{eqnarray}
as standard and the trace part
\begin{eqnarray}
\label{vector}
&f'(R)T_{i}=-\frac{3}{2}\partial_{i}f'(R)
\end{eqnarray}
which is now completely determined by the non-linearity of the $f(R)$-function.

Let us introduce now the scalar field
\begin{equation}
\label{scalaron}
\varphi=f'(F(\Sigma))
\end{equation}
related to the $f(R)$ further degrees of freedom and the effective potential
\begin{equation}
\label{potential}
V(\varphi)=\frac{1}{4}\left[\varphi F^{-1}((f')^{-1}(\varphi))+\varphi^2(f')^{-1}(\varphi)\right]
\end{equation}
which clearly disappears as soon as $f(R)=R$. Such a scalar field is used to separate the Levi--Civita contributions from the torsional ones. According to this assumption, we call $\tilde{R}_{ijhk}$ and $\tilde{\nabla}_{i}$ with $\tilde{D}_{i}$ the Riemann curvature tensor and the covariant derivatives of the torsionless Levi--Civita connection.

With these definitions, it is possible to prove that the field equations for the spin-torsion coupling \eqref{tors-matterspin} imply the antisymmetric part of the field equations for the energy-curvature coupling \eqref{curv-matterenergy}, so that only their symmetric part turns out to be significant; after having substituted \eqref{tors-matterspin} into \eqref{curv-matterenergy}, we indeed get the symmetrised Einstein-like equations in the form
\begin{eqnarray}
\label{Einstein}
\nonumber
&\tilde{R}_{ij}
=\frac{1}{\varphi^2}\left(-\frac{3}{2}\tilde{\nabla}_{i}\varphi\tilde{\nabla}_{j}\varphi
+\varphi\tilde{\nabla}_{j}\tilde{\nabla}_{i}\varphi
+\frac{1}{2}\varphi\tilde{\nabla}^{2}\varphi g_{ij}\right)+\\
\nonumber
&+\frac{i}{8\varphi}\left(\bar{\psi}\gamma_{i}\tilde{D}_{j}\psi
+\bar{\psi}\gamma_{j}\tilde{D}_{i}\psi-\tilde{D}_{j}\bar\psi\gamma_{i}\psi
-\tilde{D}_{i}\bar\psi\gamma_{j}\psi\right)+\\
&+\frac{1}{\varphi^2}V(\varphi)g_{ij}-\frac{m}{4\varphi}\bar{\psi}\psi g_{ij}
\end{eqnarray}
and, upon substitution into \eqref{matter}, we get the Dirac-like equations as in the following
\begin{eqnarray}
\label{Dirac}
&i\gamma^{k}\tilde{D}_{k}\psi-\frac{3}{16\varphi}(\bar{\psi}\gamma^{k}\psi)\gamma_{k}\psi-m\psi=0
\end{eqnarray}
as it has already been discussed in \cite{f-v}.

From now on, we shall focus only on the Dirac-like field equation \eqref{Dirac}, in which the torsional contributions are given as self-interactions of the spinor field with itself; to be more precise, the torsion-spin coupling forces the left- and right-hand projections of the spinor field to interact according to a potential of the Nambu-Jona-Lasinio type \cite{n-j--l/1,n-j--l/2}. This fact was already noticed in \cite{f}, and then generalized for the $f(R)$-gravity in \cite{f-v}; specifically, the non-linearity of the $f(R)$ function has the effect of introducing a scale factor modifying the normalization of the Dirac field, and therefore acting as a running coupling for the potential: therefore, if we have that the torsion influences the spinor dynamics by giving rise to spinorial self-interactions, we also have that the intrinsic non-linearity of $f(R)$ function has the effect of endowing these interactions with an energy-dependent coupling.

In the following, we shall exploit this approach in the case of a model of leptons, showing that the spinorial self-interactions for the two spinors will be accompanied by spinorial interactions among the two spinors that can be written in the form of electroweak interactions, featured by an energy-dependent scaling.
\section{Electroweak-like interactions from torsion}
In order to model a pair of leptons our starting point will be a coupled system of Dirac matter fields of which one is considered to be a left-handed spinor; however as we have discussed in the introduction, the double-handed spinor will be taken already massive: when in the coupled matter field equations the torsional contributions are separated away and eventually written in terms of the spin density, the fact that the spin density is the sum of the spin densities of all spinors involved implies that now there will be more interactions modelled as
\begin{eqnarray}
&i\gamma^{k}\tilde{D}_{k}e
-\frac{3}{16\varphi}\overline{e}\gamma_{k}e\gamma^{k}e
-\frac{3}{16\varphi}\overline{\nu}\gamma_{k}\nu\gamma^{k}\gamma_{5}e-me=0\\
&i\gamma^{k}\tilde{D}_{k}\nu
-\frac{3}{16\varphi}\overline{e}\gamma_{k}\gamma_{5}e\gamma^{k}\nu=0
\end{eqnarray}
where $e$ and $\nu$ denote the electron and neutrino fields.

In these equations again all the interactions are among the left- and right-hand projections, but because the neutrino has no right-hand projection then the neutrino self-interactions are lost in the neutrino field equation, leaving electron self-interactions in the electron field equation and neutrino-electron interactions in both field equations.

By employing some Fierz rearrangement, it is possible to see that the torsional potentials for the spinor fields can be transformed into the following
\begin{eqnarray}
\label{equations}
\nonumber
&i\gamma^{k}\tilde{D}_{k}e-\frac{3}{8\varphi}(\cos{\theta})^{2}\overline{e}\gamma^{k}e\gamma_{k}e
+q\tan{\theta}Z_{k}\gamma^{k}e-\\
&-\frac{g}{2\cos{\theta}}Z_{k}\gamma^{k}e_{L}+\frac{g}{\sqrt{2}}W^{*}_{k}\gamma^{k}\nu-me=0\\
&i\gamma^{k}\tilde{D}_{k}\nu+\frac{g}{2\cos{\theta}}Z_{k}\gamma^{k}\nu
+\frac{g}{\sqrt{2}}W_{k}\gamma^{k}e_{L}=0
\end{eqnarray}
once we name
\begin{eqnarray}
&Z^{k}=-\left[2(\sin{\theta})^{2}\overline{e}\gamma^{k}e-\overline{e}_{L}\gamma^{k}e_{L}
+\overline{\nu}\gamma^{k}\nu\right]\left(\frac{3\cot{\theta}}{16\varphi q}\right)
\label{neutral}\\
&W^{k}=-\left(\overline{e}_{L}\gamma^{k}\nu\right)
\left[\frac{12(\sin{\theta})^{2}-3}{16\varphi q\sqrt{2}\sin{\theta}}\right]
\label{charged}
\end{eqnarray}
where $m$ and $q$ are the mass and charge of the electron field, the parameter $\theta$ is determined by $q=g\sin{\theta}$ and the parameter $g$ is totally arbitrary; of course this parameter might have been different, but precisely because it is totally arbitrary we are free to choose it as we like, and our choice is motivated by the fact that eventually we want to recover the same set of parameters of the SM, so to have the ease of having a direct comparison.

We notice that, on the one hand, there has been a shift in the coupling constant of the electron self-interaction in the electron field equation, and on the other hand, all interactions among electron and neutrino in both field equations have been written in a form that is exactly that of the weak interactions among leptons; then we have that the mediators of such weak-like forces among leptons are composite, being them built in terms of lepton bound states whose binding is due to the presence of torsion; finally, we remark that there is no Higgs field whether composite or not appearing whatsoever: correspondingly, we notice that, firstly, as the interactions among spinors have precisely the form of the weak forces for leptons, then both this and the SM give rise to the same effective forces at the Fermi scale; secondly, since here it is in terms of leptonic bound states that the weak vector mediators are constructed while in the SM they are fundamental, then in this model we must expect the weak-like bosons to display internal structure when the energy is high enough to probe their compositeness while in the SM they must remain structureless at any energy scale, and so discrepancies at high energies must arise; finally, because the present model predicts the absence of the Higgs field, then it will be more and more confirmed as the Higgs boson is shown to be more and more elusive. The problem that now needs to be addressed is that, where the two models are supposed to give rise to the same phenomenology, there they have to yield the same predictions: we have already noticed that for both models the weak interactions for leptons have structure that are identical, and we need now to see whether their strengths can also be tuned as well.

In fact, that starting from torsion one can get weak-like forces among leptons has already been shown in \cite{F/1}, although in that paper the problem of the strength is left open; here the same results regarding the employment of torsion to get weak-like forces among leptons are obtained, but they are found, but it is in the generalized $f(R)$ gravity that they are obtained: therefore we can here exploit the presence of the field $\varphi$ as a running coupling to fit an energy-dependent scaling \cite{f-v}.
\section{The problem of gravity and electroweak interactions}
According to the above considerations, in the present approach, the weak vector bosons are the effective result of the interactions induced by torsion among lepton fields; therefore the present approach must be able to address all problems that in the SM are solved by the presence of fundamental weak bosons and Higgs boson, whether it is composite or fundamental. In the SM, the Higgs boson is necessary to give rise to the Higgs mechanism, which is a process that allows to break the gauge symmetry generating the masses of the electroweak gauge bosons; to preserve the perturbative unitarity of the S-matrix; and, finally, to preserve the renormalizability of the theory. The masses of the electroweak bosons can be written in a gauge invariant form using either the non-linear sigma model \cite{CCWZ2} or a gauge invariant formulation of the electroweak bosons. However if there is no propagating Higgs boson, quantum field amplitudes describing modes of the electroweak bosons grow too fast, violating the unitarity around TeV scales \cite{LlewellynSmith:1973ey,Lee:1977yc,Lee:1977eg,Vayonakis:1976vz}. There are several ways in which unitarity could be restored, but the SM without a Higgs boson is non-renormalizable at perturbative level. This means that if Higgs boson is not detected then a dramatic puzzle would come out in order to describe the observed masses and their hierarchies of the particles.

A way out could be that the electroweak interactions may become strongly coupled at the TeV scales and then the related gauge theory becomes unitary at non-perturbative level. Yet another possibility for models without a Higgs boson may consist in introducing weakly coupled new particles to delay the unitarity problem into the multi TeV regime where the ultraviolet limit of the SM is expected to become relevant. These ideas are very intriguing and show several features of electroweak interactions: first of all, the Higgs mechanism is strictly necessary to generate masses for the electroweak bosons if these electroweak boson are initially fundamental gauge fields; beside, some mechanisms can be unitary but not renormalizable or vice-versa. In summary, the paradigm is that three different criteria should be fulfilled: $i)$ the masses of electroweak bosons must be generated if they initially are gauge fields or be always present if they are composite fields; $ii)$ the perturbative unitarity must be respected; $iii)$ the renormalizability of the theory must be ensured at any scale of energy.

In any case, it is possible to define an action in terms of an expansion in the scale of the electroweak interactions $v$: the action can be written as \cite{calm10}
\begin{eqnarray}
\label{effaction}
\mathscr{L}_{\rm{SM}}=\mathscr{L}_{\rm{SM \ Higgsless}}+\sum_i \frac{C_i}{v^N} O^{4+N}_i\,,
\end{eqnarray}
where $O^{4+N}_i$ are operators compatible with the symmetries of the model.

The analogy between the effective action for the electroweak interactions (\ref{effaction}) and
 $f(R)$ gravity is striking assuming a gravitational Taylor series of the form
\begin{equation}
\label{effaction2}
\mathscr{L}_{f(R)-\rm{gravity}}=f(R)\equiv\Lambda+\sum_k \frac{1}{k!}f_{0}^{(k)}R^{k}\,,
\end{equation}
where the coefficients $f_{0}^{(k)}$ are the $k$-order derivatives of $f(R)$ at a certain value of $R$; it is straightforward that the extra gravitational degrees of freedom can be suitably transformed into the above scalar field $\varphi$ which allows to avoid the hierarchy problem. It is clear that both the electroweak interaction and $f(R)$ gravity have a dimensional energy scale: the Planck mass sets the strength of gravitational interactions while a given weak scale $\lambda$ determines the range and the strength of the electroweak interactions. If gravity shows a running coupling, these scales can be compared at TeV energies and then probed at LHC. If this is done, the electroweak-like interactions would have the right strength, beside the already noticed correct structure, of the known electroweak interactions, and the electroweak bosons would not be the standard gauge bosons but they would be recovered from the further gravitational degrees of freedom coming from torsion in $f(R)$ gravity. In this case however, the masses cannot be generated by a mechanism of mass generation but they will have to be always present; actually this is precisely what happens in the present construction.

Specifically, in the Standard Model, the $W$ and $Z$ bosons are the result of a spontaneous symmetry breaking being originally gauge fields. In our case, they are not gauge fields from the beginning but they naturally arise from torsional interactions among leptons and do not come from any symmetry breaking. However, in a more general approach, as that discussed in \cite{CBD}, $W$ and $Z$ bosons, as well as $SU(2)_L$ and $SU(3)$ interactions, can be related to a gravitationally induced symmetry breaking connected to a dimensional reduction.

In the present case, by considering the expressions for the weak-like bosons (\ref{neutral}-\ref{charged}) we have the following equations
\begin{eqnarray}
&\nabla_{\mu}Z^{\mu}+\left(\frac{3\cot{\theta}}{16}\right)
\frac{m}{q}\frac{i\overline{e}\gamma_{5}e}{\varphi}+Z^{\mu}\partial_{\mu}\ln{\varphi}=0
\label{conservedneutral}\\
&\nabla_{\mu}W^{\mu}-\left[\frac{12(\sin{\theta})^{2}-3}{16\sqrt{2}\sin{\theta}}\right]
\frac{m}{q}\frac{i\overline{e}\gamma_{5}\nu}{\varphi}+W^{\mu}\partial_{\mu}\ln{\varphi}=0
\label{conservedcharged}
\end{eqnarray}
showing that the vector bosons are indeed massive as they satisfy partially conserved axial currents, and the problem related to the masses of the vector boson is addressed. These expressions, given in terms of $\varphi$ as obtained here, and the corresponding expressions in the SM, given in terms of the Higgs field, as shown for instance \cite{F}, can be compared, and the partially conserved axial currents display in both cases astonishing similarities.

A gravitational action like (\ref{effaction2}) is in principle non-perturbatively renormalizable if, as shown by Weinberg, there is a non-trivial fixed point which makes the gravity asymptotically free \cite{fixedpoint}; this scenario implies that only a finite number of Wilson coefficients in the effective action would need to be measured and the theory would thus be predictive and probed at LHC. Measuring the strength of the electroweak interactions in the electroweak $W$-boson scattering could easily reveal a non-trivial running of the electroweak scale $v$; if such an electroweak fixed point exists, an increase in the strength of the electroweak interactions could be found, as in the strongly interacting W-bosons scenario, before the electroweak interactions become very weak and eventually irrelevant in the fixed point regime. In analogy to the non-perturbative running of the non-perturbative Planck mass, it is possible to introduce an effective weak scale
\begin{eqnarray}
v_{\rm{eff}}^2=v^2\left(1+\frac{\omega}{8\pi} \frac{\mu^2}{v^2} \right)\,,
\end{eqnarray}
where $\mu$ is an arbitrary mass scale, $\omega$ a non-perturbative parameter which determines the running of the effective weak scale and $v$ is the weak scale measured at low energies: if $\omega$ is positive, the electroweak interactions would become weaker with increasing center of mass energy, and this asymptotically free weak interaction would be renormalizable at non-perturbative level without the need of a Higgs boson, solving the issue raised above to preserve the perturbative unitarity of the S-matrix.
Besides, asymptotically free weak interactions induced by gravity could solve the unitarity problem of SM (see e.g. \cite{calm10}).

In summary, problems like unitarity, renormalization and mass generation, could be, in principle, addressed by a model even without the Higgs boson, if specific dynamics, settling a suitable running coupling is found; as we will see below, such a dynamics can be implemented by the function $\varphi$.
\section{Energy-dependent coupling induced by $f(R)$-gravity}
Let us assume now the action \eqref{effaction2}. Specifically, the $f(R)$ function is assumed analytic, the least-order is the Einstein action; the second-order does not contribute to the energy trace condition, and therefore we will consider the third-order power, truncating all the following higher-order powers. The Taylor coefficients can be written as
\begin{eqnarray}
\label{formf}
&f(R)=R+\frac{1}{4}\varepsilon R^{2}+\frac{1}{27}\eta^{2}\varepsilon^{2}R^{3}
\end{eqnarray}
in terms of the two parameters $\varepsilon$ and $\eta$ that have to be determined eventually. The $\Lambda$ term has been discarded at the moment since it is unessential for the following considerations.

In this situation, equation \eqref{trace} is a cubic equation in $R$ with three solutions, of which we are going to take the only solution that is always real; this is also the only solution for which $R$ vanishes as the energy trace $\Sigma$ vanishes: with it we have that $\varphi$ is
\begin{eqnarray}
\label{formvarphi}
\nonumber
&\varphi=3\left(1-\frac{3}{8\eta^{2}}\right)
+\left(\frac{3}{4\eta}+\sqrt[3]{s+\sqrt[2]{s^{2}-1}}\right)^{2}+\\
&+\left(\frac{3}{4\eta}+\sqrt[3]{s-\sqrt[2]{s^{2}-1}}\right)^{2}
\end{eqnarray}
function of the energy $\frac{1}{2}\eta\varepsilon\Sigma=s$ in the parameter $\eta$ alone.

Therefore in the torsionally induced electroweak interactions, we have that the coupling $\frac{1}{\varphi}$ is energy dependent and expressed as a scale factor in one parameter; as an easy analysis shows, this coupling starts from the unity and then it increases up to its maximum value before eventually decreasing to vanish asymptotically as the energy starting from zero increases to infinity. Henceforth the leptons would start with negligible interactions but then they would get larger scattering amplitudes before finally being asymptotically free.

An important remark is in order at this point. The form of action \eqref{formf} is extremely relevant since it can be shown that it passes several viability conditions as an alternative gravitational theory to General Relativity. In \cite{stabile}, it is shown that a large class of analytic $f(R)$ models can be recast in this way leading to self-consistent Newtonian and post-Newtonian limits of the theory. This means that the coefficients of the expansion are the leading parameters setting the scale of the interaction. Examples of stable self-gravitating systems constructed according to an action of the form \eqref{formf} are discussed in \cite{salzano,napolitano}. Cosmological tests according to such an action are discussed in \cite{cosmography}.

In the following we shall discuss some phenomenological consequences of this approach at electroweak scales.
\section{Probing gravity at TeV scales}
The results we have found show that torsion induces between a couple of leptons interactions whose structure is identical to that of the weak forces, while the mass generation of the resulting composite weak vector boson in unitary-renormalizable Higgsless models is addressed within the frame of $f(R)$ gravity with an energy-dependent scale by exploiting the $\varphi$ function as a running coupling: these theoretical results are extremely intriguing in themselves, and especially if we think that they all could be investigated in the range between GeV and TeV scales at LHC by experiments such as ATLAS and CMS.

It is important to stress that any ultraviolet model of gravity (e.g. at TeV scales) have to explain also the observed weakness of gravitational effects at largest (infrared) scales; this means that massless (or quasi-massless) modes have to be considered in any case so that the results of standard General Relativity are reproduced in the low energy regime \cite{greci}. What this implies is that the effective gravitational energy scale (Planck scale) has to be rescaled according to $\frac{1}{\varphi}$. In terms of the mass parameter, being $M_P^2=\frac{c\hbar}{G_N}$ the constraint coming from the ultraviolet limit of the theory ($10^{19}$ GeV), we can generically set $ M_{\rm{eff}}^2 =M_P^2 G(\varphi)$, where $M_{\rm{eff}}$ is a cut-off mass that becomes relevant as soon as the Lorentz invariance is violated and $G(\varphi)$ is a function of $\varphi$ that has to be determined. Such a scale in the context discussed here could be at TeV scales. As shown above, it is quite natural to have effective theories containing scalar fields of gravitational origin; in this sense $M_{\rm{eff}}$ would result as a running coupling. To be more explicit, the dynamics is led by the above effective potential $V(\varphi)$ in equation \eqref{potential} and the non-minimal coupling $f'(\varphi)$; such functions could be experimentally tested since they are related to massive states: in particular, the effective potential can be phenomenologically chosen to be
\begin{eqnarray}
\label{Higgs}
V(\varphi)=\frac{M_{\rm{eff}}^2}{2} \varphi^2-\frac{\lambda^{2}}{4} \varphi^4\,,
\end{eqnarray}
from which it is easy to derive the vacuum expectation value of $\varphi$ as the fundamental scale of the theory that has to be eventually compared to the Higgs vacuum expectation value measured to be 246 GeV. This is the standard choice of quantum field theory which perfectly fits the above arguments, although here the choice is not {\it ad hoc} because the scalar field $\varphi$ is not put {\it by hand} into dynamics, being it given by the extra degrees of freedom of gravitational field that are naturally present in the general set-up of the $f(R)$ gravity.

To be more precise, the form of the effective potential
\eqref{Higgs} can be reconstructed from the conformal potential
\begin{eqnarray}
\label{h9}
\frac{V(\varphi)}{\varphi^2} = -\frac{1}{2}\frac{f(R) - f'(R)R}{f'(R)}\,,
\end{eqnarray}
which gives Eq. \eqref{potential} starting from the position
\eqref{scalaron}. The effective mass $M_{\rm{eff}}$ and the
self-interaction parameter $\lambda$ are related to the effective
Klein-Gordon equation that comes out from the conformal
transformation of $f(R)$ gravity. For details see \cite {CBD} and
\cite{CCSV1}.

Let us discuss now a couple of features of the model. First of all, we have the {\it hierarchy problem}. It is important to recall that the hierarchy problem occurs when couplings and masses of effective theories are very different from the parameters measured by experiments, happening because the measured parameters are related to the fundamental parameters by renormalization so that cancellations between fundamental quantities and quantum corrections are necessary. The hierarchy problem is essentially a fine-tuning problem: if $M_{\rm{eff}}$ is larger than Higgs mass, then the hierarchy problem is circumvented. A second feature concerns the {\it relative strengths} of weak and gravitational forces. As it has been noticed in \cite{h-h-k-n}, in units $\hbar=c=1$ both electromagnetic and strong interactions have dimensionless couplings while both weak and gravitational interactions have couplings with dimensions of a length; although this does point toward the fact that unifications between weak and gravitational forces may be possible, nevertheless any of these unifications shall be possible only after that the two fundamental lengths given by $l_{\rm{weak}}\sim 10^{-18} \rm{m}$ and $l_{\rm{gravity}}\sim 10^{-34} \rm{m}$ have been set to the same value, that is the Fermi constant for the weak force and the Newton constant for gravity must be equal. However the weak force is much stronger than gravity, unless a cancellation between the bare value of Fermi constant and its quantum corrections occurs, or alternatively the Higgs boson is much lighter than the Planck mass, unless what occurs is a fine-tuning between the quadratic radiative corrections and the Higgs bare mass; present data suggest the Higgs mass should be between 115 GeV and 350 GeV with different selected decay channels from $b\bar{b}$ to $t\bar{t}$ \cite{camy}, but with this state of the art, the problem cannot be formulated in the context of SM, where the Higgs mass cannot be calculated, and this problem would be solvable only if, in a given effective theory of particles in which the Higgs boson mass can be calculated, no fine-tuning is present. If one accepts the {\it big-desert} assumption and the existence of a {\it hierarchy problem}, then some new mechanism at Higgs scale becomes necessary to avoid fine-tuning. The model we are discussing contains a running coupling that allows us not to set the Higgs scale: if the mass of the field $\varphi$ is in the TeV region, there is none of the above problems, being $\varphi$ a gravitational scale; in this case, the SM holds up plus an extended gravitational sector with torsion. In other words, the Planck scale can be dynamically derived from the vacuum expectation value of the function $\varphi$; such a scale can be recovered, as soon as the coupling $\lambda$ is of the order of $10^{-31}$, and our $f(R)$-model with torsion is valid up to the cut-off scale $M_{\rm{eff}}\sim$ TeV. The tiny value of $\lambda$ allows the presence of physical (quasi-) massless gravitons with very large interaction lengths. It is important to stress that, via a conformal transformation from the Jordan frame to the Einstein frame, the Planck scale is decoupled from the vacuum expectation of the scalar field $\varphi$; on the other hand however, the scalar field redefinition has to preserve the vacuum of the underlying background. Besides, the gauge couplings and masses depend on the vacuum expectation value of $\varphi$ and they are dynamically determined. This means that both the SM and Einstein Gravity (in the above conformal-affine sense) could be recovered without the {\it hierarchy problem}.

To conclude, we wish to go back to the problem of mass generation. In the present model we have described the vector bosons as composite fermionic bound states; fermion scattering producing bound states has also been discussed in \cite{Antoniadis:2001sw, Antoniadis:1998ig}, while compositeness was discussed in \cite{Meade:2007sz}: assuming that the particles of the SM have sizes that are related to the cut-off, then the vector bosons would have corresponding ranges related to the same cut-off, that is of the order of $M_{\rm{eff}}^{-1}$. Potentially, the formation of bound states could mimic the decay of semi-classical quantum black holes and, at lower energies, it could be useful to investigate substructures of the SM. Strong scattering effects could emerge in the TeV region involving the field $\varphi$ coupled to the SM fields. Bounds on the production of mini-black holes can be derived from astroparticle physics \cite{Feng:2001ib, Anchordoqui:2001cg, Anchordoqui:2003jr, Ringwald:2001vk, Kowalski:2002gb}, where in particular in \cite{Anchordoqui:2001cg}, a bound on the cross-section is given as
\begin{eqnarray}
\sigma_{\nu N \to BH + X} < \frac{0.5}{\mbox{TeV}^2}.
\end{eqnarray}
The cross-section $\sigma \simeq M_{\rm{eff}}^{-2}$ gives a bound at TeV scales: this means that strong scattering processes at LHC would have the cross-section of the order of magnitude of
\begin{eqnarray}
\sigma_{pp \to \rm{grav. modes}+X}\sim 1 \times 10^7 \mbox{fb}\,,
\end{eqnarray}
which dominates the cross-sections expected for the SM. In this case, the hierarchy problem would not be present.
\section{Discussion and Conclusions}
In this paper, we have considered $f(R)$-gravity with torsion in relation to electroweak interactions. The presence of torsion tensor gives rise to interactions that have the structure of the weak interactions among the leptons and the induced coupling results energy-dependent. Assuming analytical $f(R)$-functions, it is possible to show that the coupling is running with initial weak strength able to further increase before vanishing in the end. This phenomenology gives rise, for interacting fermions, to a scattering amplitude that, even if it is low at the beginning can become larger before vanishing: as it is widely known, torsion is very weak in low-energy and non-renormalizable in high-energy systems, and therefore these torsionally induced electroweak-like interaction would correspondingly be very weak in the infrared and non-renormalizable in the ultraviolet regime; nevertheless, in $f(R)$ models, the coupling can become relevant even for large-scale before vanishing at the short-scale physics, and henceforth the torsionally induced electroweak-like interaction in $f(R)$-gravity can become relevant at the Higgs scale before becoming renormalizable at the Planck scale. The approach based on the presence of torsion in $f(R)$-gravity gives the possibility to have torsion both relevant before the Planck scale and negligible at the Planck scale, achieving a double-take only in terms of the single assumption of being in a specific $f(R)$ model. In this way, torsion gives the possibility to induce electroweak-like interactions bypassing the Higgs mechanism and the hierarchy problem. Therefore $f(R)$-gravity could ensure the possibility to properly fit the scale of specific interactions whose form is given by torsion to be that of the weak forces.

This is, we believe, important in view of the solution of the problem of unification in physics: in fact, as our analysis has shown, it is possible to employ a specific $f(R)$ model to fit the running coupling of some interactions and to use torsion to give rise to those interactions; hence, interactions like the weak forces are, both in strength and in structure, entirely derived within the context of gravitational interactions. Furthermore, in this context, there is no need to postulate any new particle, since the non-linearity of the $f(R)$ function comes from the requirement that the action of gravitational field is not restricted to be the simplest Einstein-Hilbert action; on the other hand, the presence of torsion comes from the requirement of a connection that is not restricted to be the simplest Levi--Civita symmetric connection. If the action has to be written in terms of the Ricci scalar curvature, then $f(R)$-gravity is the most general action that we may consider; besides, if the covariant derivatives have to be given in terms of metric-compatible connections, then connections with torsion are the most general one we may have. The further geometric degrees of freedom, not present in general relativity, naturally acquire the role of a scalar field enriching the dynamics \cite{book}.

In this sense, the electroweak interactions may be the effect of the presence of torsion in $f(R)$-gravity, a theory that is the most general and straightforward geometric background for any model we want to develop upon it. The philosophical implications of this approach are profound, as this would force us to rethink the issue of unification in physics, an issue that is still open and for which no advance has been obtained since the when the decay of the proton, foreseen by the grand unification and promised by the $SU(5)$-model, failed to be discovered. On the other hand, the results that we have presented here seem to fit very well the behavior of the weak interactions between GeV and TeV scales. This fact does not imply that such fitness will be maintained at any scale, and on the contrary, discrepancies between this and the SM could appear beyond the TeV scale, making our approach hopefully more predictive than the SM itself.

As the present generation of accelerators is already running up toward those energies we think that a very interesting moment for physics is coming, and the approach we have presented here could fuel the debate in a different but not less important direction.

\end{document}